# Cascade at local yield strain for silica and metallic glass

Nandlal Pingua,[1,2] Himani Rautela,[2] Roni Chatterjee,[3] Smarajit Karmakar,[3] Pinaki Chaudhuri,[4,5] and Shiladitya Sengupta*[2]
[1)] *Dept. of Chemical Engineering, National Institute of Technology, Tiruchirappalli, Tamil Nadu 620015, India.*
[2)] *Dept. of Physics, Indian Institute of Technology Roorkee, 247667 Uttarakhand, India.*
[3)] *Tata Institute of Fundamental Research, 36/P, Gopanpally Village, Serilingampally Mandal,Ranga Reddy District, Hyderabad, 500046, Telangana, India.*
[4)] *The Institute of Mathematical Sciences, CIT Campus, Taramani, Chennai 600113, India.*
[5)] *Homi Bhabha National Institute, Anushaktinagar, Mumbai 400094, India.*

We report observations of unusal *first* plastic events in silica and metallic glasses in the shear startup regime at applied strain two orders of magnitude smaller than yield strain. The (non-Affine) particle displacement field during these events have complex real space structure with multiple disconnected cores of high displacement appearing at the *same* applied strain under athermal quasistatic simple shear deformation, and identified by a "cell based cluster analysis" method. By monitoring the stress relaxation during the first plastic event by Langevin dynamics simulation, we directly show the cascade nature of these events. Thus these first plastic events are reminiscent of avalanches in the post-yielding steady state, but unlike the steady state avalanches, we show that these events are not system spanning. To understand the nature of these events, we tune three factors that are known to affect brittleness of a glass. These are (i) sample preparation history, (ii) inter-particle interactions and (iii) rigidity of the background matrix applying a "soft matrix" probe recently developed by some of us. In each case we show that such first plastic events are more probable in more ductile glasses. Our observations are consistent with the picture that more ductile materials are softer, implying that understanding the role of softness may be a promising route to develop microscopic quantifiers of brittleness and thus clarifying the physical origin of brittle-to-ductile transition.

## I. INTRODUCTION

Glasses are ubiquitous amorphous materials with enormous engineering applications[1]. Yet, many physical properties, such as the shear deformation response remains poorly understood[2,3]. Nevertheless, research over the last several decades have shown that the shear deformation in glasses proceeds *via* a sequence of sudden, irreversible particle rearrangement events called "shear transformations" or "plastic events" or "avalanches"[4–7]. Consequently, the stress-strain curves of typical glasses are "serrated" with plastic discontinuties found even at small deformation in the elastic (Hookean) regime[4,8,9].

The most well-known mechanical response of a (structural) glass is perhaps its brittleness and catastrophic failure under deformation. This is quantified by the fact that structural glasses have comparable shear modulus and higher tensile strength but orders of magnitude lower fracture toughness than typical ductile solids (*e.g.* crystalline metals)[10,11]. The brittle-like response is further characterised by a sharp stress drop at the yielding transition and concomittant development of signature strain localization known as shear bands[10] which eventually leads to the yielding transition[3]. Thus we note that it is possible to distinguish the brittle and ductile-like responses of glasses at the *macro*-scale, at least qualitatively[12]. However, at the *micro*-scale, there are fundamental conceptual challenges in quantifying disorder and predicting local mechanical instability in glasses[1,3,13]. Consequently, how the macro-scale differences between brittleness and ductility emerges from micro-scale physical processes, is not very clear and a central question of current research[3].

Particle-based microscopic simulations - especially in the athermal, quasi-static (AQS) limit where the thermal noise is absent and relaxation times are infinite - have played a major role in establishing the phenomenology of shear deformation response of amorphous materials[4,8,9,14–18]. In the AQS limit, the plastic events are typically detected by jump discontinuities in the stress and the potential energy of the system. As the applied strain increases, successive events become correlated in space and time leading to cascade, shear band formation and eventually yielding transition[4]. Scaling analyses of the average size of the events have revealed a power law exponent and a system-size dependent cutoff[8,19–23]. Such analyses have mostly been carried out in the post-yielding steady state where avalanches are system-spanning[4]. However, recently several studies have focussed on the initial, transient, elastic shear-startup regime at *small* applied strain - typically 0-4% - much below the yield strain[20–22,25–27]. One motivation came from the puzzle of whether the avalanche exponent changes with applied strain[22,24,25,27]. Another line of research stems from the recent prediction that the potential energy landscape (PEL) of the system has a complex, hierarchial structure such that the initial undeformed sample is "marginally stable", *i.e.*, transition between basins in the PEL in the limit of zero barrier height is possible[28,29]. Later Ref.[20] made a more detailed prediction about avalanches due to marginal sta-bility in a mean-field model with further support from simulations[21].

Along with avalanche statistics and topographical analysis of the PEL, description of stress and strain fields and their inter-relationship in real space is essential to get a comprehensive understanding of plastic events. An important question here is how successive plastic events in real space are triggered, *i.e.* the spatial (and temporal) correlation among plastic events. The correlator of the stress field due to a plastic event is given by the so-called Eshelby solution, with long range decay and quadrupolar symmetry[16,30]. However, owing to the amorphous structure of glasses, the (plastic component of the) strain field and thus the stress field at any given deformation state are *a priori* unknown, making the question difficult to answer. In this regard, we note the following points. First, most studies of real space aspect of avalanches have been done in the steady state and only a few relatively recent studies have investigated the initial elastic start-up regime[31]. The conventional wisdom is that when the applied strain is close to zero, the stress response to an applied strain should be dominated by the elasticity of the medium and plastic events should be isolated, with the core of the event containing $\sim 10-100$ atoms and localized in space and time[32,33]. This picture makes sense intuitively, because the energy (stress) released due to a plastic event should be small and barrier for rearrangement should be large in the elastic regime. However, if the initial undeformed sample is indeed marginally stable, a more complex real-space structure of the event is an intriguing possibility. Second, most computational studies of shear deformation response use soft spheres interacting *via* Lennard-Jone (LJ) type short-ranged, isotropic interactions as model glass. For engineering applications, it is necessary to consider more realistic materials such as oxides and metallic glasses, which are however, relatively less explored in the elastic regime. These materials have more complex and longer-ranged inter-particle interactions than LJ-type glasses, which may also lead to more complex correlation in the strain field. Indeed, a recent study have found that in metallic glasses even for the first few plastic events, an event may perturb the elastic background matrix triggering the next event[26].

Here we study a proto-type oxide glass - silica as well as a model metallic glass - $Cu_{64.5}\,Zr_{35.5}$ and focus mainly on the *first* plastic event, in the so-called elastic shear start up regime dominated by linear or piecewise linear elastic response[34] and at applied strain typically two orders of magnitude less than the yield strain. We report that in complex materials such as silica and metallic glass, even the *first* plastic event can exhibit complex cascade type nature. Although this may be counter-intuitive, we show in subsequent sections that this is a manifestation of silica and metallic glass being relatively *more* ductile than LJ type materials.

The paper is organized as follows: we briefly describe the model glasses and simulation details in Sec. II B. Then we present the results in Sec. III. First in Sec. III A, we establish the robustness and reliability of the observation of *first* plastic events with multiple distinct cores of high non-Affine displacement. Then we swtich to Langevin dynamics to monitor stress relaxation dynamics and reveal the cascade nature of the events in Sec. III B. Next we modify brittleness of the material in three different ways: (i) by changing the degree of annealing of the initial undeformed sample Sec. III C, (ii) by changing the inter-atomic potential III D and (iii) by tuning the rigidity of the background matrix using a new local probe of mechanical response termed as the “soft matrix” (SM) method (Sec. III E), and study the effect on the probability of the observed events. Finally we summarize the results and discuss the conclusions in Sec. IV.

## II. SIMULATION DETAILS

### A. Models

We study two different rigid-ion type empirical force-fields of silica[35], an embedded atom model of metallic glass as well as a long ranged LJ-type model in the present work:

*a.* ***VSPBKS silica:*** It is a modified version of the Beest-Karmer-Santen (BKS) model[36] introduced by Saika-Voivod *et al.*[37], and denoted here as the VSPBKS model. The interaction potential of the VSPBKS model is *entirely two-body* of the form:

$$
\begin{aligned}
U(\vec{r}^N) &= \sum_{<ij>} \phi_{ij}^{VSPBKS} \\
\phi_{ij}^{VSPBKS} &= \phi_{ij}^{BKS} + \left[4\epsilon_{ij}\left(\frac{\sigma_{ij}}{r_{ij}}\right)^{30} - 4\epsilon_{ij}\left(\frac{\sigma_{ij}}{r_{ij}}\right)^{6}\right] \\
\phi_{ij}^{BKS} &= A_{ij}e^{-B_{ij}r_{ij}} - \frac{C_{ij}}{r_{ij}^6} + \frac{q_i q_j}{4\pi\epsilon\, r_{ij}}
\end{aligned}
\qquad (1)
$$

where $r_{ij}$ is the inter-atomic separation between distinct pairs $i$ and $j$, $A_{ij}, B_{ij}, C_{ij}$ are constants and $q_i, q_j$ are particle charges. The original BKS model describes short-raged steric repulsion, van der Waals attraction and Coulomb interaction between charged silicon and oxygen particles. The modfication done by Ref.[37] is to ensure that the potential remains repulsive in the limit of vanishing inter-particle distance. Ref.[37] implemented a soft cutoff at $r_s \approx 7.75\mathring{A}$ and a hard cutoff at $r_c = 10\mathring{A}$.

*b.* ***Vashishta silica:*** This force-field, denoted here as the Vashishta model[38,39], includes a 3-body interaction in addition to the 2-body interactions, to *explicitly* describe covalent bonding of the silica tetrahedral net-work. The interaction potential is of the form

$$U(\vec{r}^N) = \sum_{<ij>} \phi^{(2)}_{ij} + \sum_{<ijk>} \phi^{(3)}_{ijk}$$
$$\phi^{(2)}_{ij} = \frac{H_{ij}}{r_{ij}^{\eta_{ij}}} - \frac{D_{ij}}{r^4_{ij}} e^{-r_{ij}/\lambda_{4,ij}} + \frac{q_i q_j}{r_{ij}} e^{-r_{ij}/\lambda_{1,ij}}$$
$$\phi^{(3)}_{ijk} = \frac{B_{ijk}(cos\theta_{ijk} - cos\theta_0)^2}{1 + C_{ijk}(cos\theta_{ijk} - cos\theta_0)^2} e^{\frac{\gamma_{ij}}{r_{ij}-r_0} + \frac{\gamma_{ik}}{r_{ik}-r_0}} \quad (2)$$

where, $\phi^{(2)}_{ij}$ and $\phi^{(3)}_{ijk}$ are the 2- and the 3-body interaction terms respectively. The 2-body interaction describes short-ranged steric repulsion of strength $H_{ij}$ and steepness exponent $\eta_{ij}$, attractive charge-dipole interaction of strength $D_{ij} = \frac{\alpha_i q_j^2 + \alpha_j q_i^2}{2}$ and (screened) Coulomb interaction between silicon and oxygen point charges. Here $q_i, \alpha_i$ are respectively the electric charge and electronic polarizability of the $i$th ion, and $\lambda_{1,ij}, \lambda_{4,ij}$ are screening lengths for the Coulomb and the charge-dipole interactions respectively. The 3-body potential accounts for cost of bond energy with a preferential bond angle $\theta_0$ made by a triplet of particles $< ijk >$, set to that of a tetrahedral network[38]. Interaction cutoff of the 2-body and 3-body terms are set to $r_c = 10\mathring{A}$ and $r_0 = 2.6\mathring{A}$ respectively and the two body potential is truncated and shifted such that both the two body potential and its first derivative smoothly goes to zero at $r_c$[39].

*c.* ***Metallic glass:*** We study an embedded atom model of binary metallic glass alloy of Cu-Zr with ratio $Cu : Zr = 64.5 : 35.5$. Details of the model are described in Ref.[40].

*d.* ***Long-ranged Kob-Andersen (LKA) model:*** The Kob-Andersen (KA) model is a prototype glass where binary mixture of soft spheres interact *via* Lennard-Jones type interactions. The potential is given by

$$U(\vec{r}^N) = \sum_{<ij>} \phi_{ij}$$
$$\phi_{ij} = 4\epsilon_{ij} \left[ \left(\frac{\sigma_{ij}}{r_{ij}}\right)^{12} - \left(\frac{\sigma_{ij}}{r_{ij}}\right)^{6} + C_0 + C_2 \left(\frac{r_{ij}}{\sigma_{ij}}\right)^{2} \right] \quad (3)$$

where $i, j$ are particle indices and there are two species $A, B$. In case of the long-range Kob-Andersen model the interaction cutoff is set at $r_c = 4.5\sigma_{ij}$ making the interaction range longer than the standard value. The constants $C_0, C_2$ are chosen so that $\phi_{ij}$ are smooth upto the second derivative at $r_c$. Like the standard KA model, the composition is $A : B = 80 : 20$, ratio of diameters $\sigma_{AB}/\sigma_{AA} = 0.8, \sigma_{BB}/\sigma_{AA} = 0.88$ and ratio of interaction strength $\epsilon_{AB}/\epsilon_{AA} = 1.5, \epsilon_{BB}/\epsilon_{AA} = 0.5$. Here $\sigma_{AA}$ and $\epsilon_{AA}$ are units of length and energy respectively.

## B. Methods

*a.* ***Sample preparation protocol:*** For all the models, we perform large-scale, particle-based molecular dynamics (MD) simulations using LAMMPS[23,41,50]. For both VSPBKS and Vashisht silica, the NVT ensemble is used at a chosen density $\rho$=2.8 gm/cm$^3$ and a parent temperature $T_p$ to generate equilibrium trajectories. Velocity-Verlet algorithm is used to integrate the equation of motion with an integration time step of 1 fs. The temperature is controlled by Nosé-Hoover thermostat with relaxation times of 100 fs. For the VSPBKS silica, seven equilibrated MD trajectories are generated at $T_p = 3078\,K$ with runlength of 13107200 fs for each. The system sizes used are $N = 13824, 46656, 216000$ particles, with most analyses done using $N = 46656$. For the Vashishta silica, one equlibrium trajectory of runlength 6553600 fs at $T_p = 5000\,K$ is generated for a system size $N = 46656$. We have simulated the $Cu_{64.5}Zr_{35.5}$ metallic glass with system sizes $N = 50000, 10^5, 150000$ in a cubic lattice maintaining periodic boundary conditions. Equilibrium MD trajectories are generated at $T_p = 1300\,K$ at zero pressure using the NPT ensemble and are then cooled to $T = 300K$ at a rate of $10\,K/ps$ to make room temperature glasses. The long-range Kob-Andersen model is equilibrated NVT MD for a system size $N = 50000$ at number density $\rho = 1.2$ and temperature $T = 1.0$ (reduced units). We prepared 30 independent samples for this case.

Next, *quenched* configurations at mechanical equilibrium are prepared from equilibrated MD configurations (for silica and LKA) or from room temperature glass (for metallic glass) by minimizing the potential energy on LAMMPS using standard algorithms. We use two different minimization algorithms - the conjugate-gradient (CG) and the steepest descent (SD) in the present analysis. These configurations are used as initial samples for yielding analysis.

*b.* ***Shear deformation protocol:*** We apply uniform athermal quasistatic shear in XY plane to all particles (Affine deformation to particle $i$: $x_i^{'} = x_i + \delta\gamma\, y_i,\ y_i^{'} = y_i,\ z_i^{'} = z_i$), followed by potential energy minimization of the resultant configuration. For both the silica models, the default strain increment is $\delta\gamma = 5 \times 10^{-5} = 0.005\%$ (unless specified otherwise) and the maximum strain is $\gamma_{max} = 2\%$. For the metallic glass, the default $\delta\gamma = 10^{-6} = 0.0001\%$ and $\gamma_{max} = 0.5\%$.

*c.* ***Identification of a plastic event:*** To detect a plastic event and compute the local yield strain $\gamma_p$, we follow Ref.[33] and monitor several indicators, *viz.*, discontinuities in per particle potential energy, stress (XY component) and the mean squared displacement MSD (averaged over particles, but not time-origin), and a newly defined indicator called the "local maximum displacement" (LMD): $\Delta z^2_{max}(\gamma) = \text{maximum}[z_i(\gamma+\delta\gamma) - z_i(\gamma)]^2$ where $z_i(\gamma)$ is the coordinate of particle $i$ perpendicular to the shear direction. We have found the LMD to be a computationally cheap but robust indicator of plastic events[33].

Here we monitor the LMD and identify the (first) plastic event when this indicator (first) exceeds a threshold value. More details are provided in appendix A.

*d.* ***Langevin dynamics simulation:*** After detecting the first plastic event by AQS, NV Langevin dynamics (LD) is performed using LAMMPS, starting from the deformed configuration at the local yield strain $\gamma_p$, to study the time evolution of stress relaxation. We analyzed both the VSPBKS silica and the metallic glass models by LD. The velocity-verlet algorithm was employed to integrate the equation of motion with a time step of 1 fs. During LD, the start and the end temperatures are set to be zero. The damping parameter is set in the range 0.6-1.0 for the VSPBKS silica and to 10 for the metallic glass. 5 samples of "isolated" and 5 samples of "multiple" -type first plastic events are monitored for VSPBKS silica for system size $N = 46656$ particles. For the metallic glass, LD is studied over 2 "isolated" type and 3 "multiple" type samples for system size $N = 50000$.

## III. RESULTS

### A. Observation in real space of "non-isolated" first plastic events

When an amorphous solid is deformed by applied shear, particles undergo sudden, irreversible, non-Affine rearrangement to release accumulated stress. While it is known that for large strain approaching yielding point, these so-called plastic events trigger one another to form system-spanning cascade[3,4], at small deformation in the elastic regime, the *core* of the event (non-Affine displacement field) are expected to be isolated and localized in space. We estimate the non-Affine part of the displacement field by computing the incremental squared displacement (SD) perpendicular to the applied shear for every particle at the first plastic event occurring at $\gamma_p$ ("local yield strain"). Surprisingly, in silica and metallic glass, we observe that even for the *first* plastic event, the real space structure of the displacement field can be distinctly non-localized with *multiple, disconnected core regions* of high displacements appearing at the *same* applied strain $\gamma_p$. Visualizations of the displacement fields are shown in Fig. 1 for poorly annealed samples of VSPBKS silica for two different system sizes, $N = 46656$ (panels a,b) and $N = 216000$ (panels c,d), and for metallic glass (panels e,f). For clarity, particles whose $z$ component of SD in a single strain step increment $\delta\gamma$ exceeds a threshold are highlighted. Multiple disjointed regions with high displacements are clearly visible, and is more prominent for the higher system size. Note that at such small deformation ($\gamma_p \sim 0.1-0.3\%$ for both silica and metallic glass), non-isolated non-Affine displacement fields are not impossible but are considered to be highly improbable in glasses with Lennard-Jones (LJ) type interactions. To the contrary, for the VSPBKS silica out of 134 samples, 84 shows such non-local first plastic events.

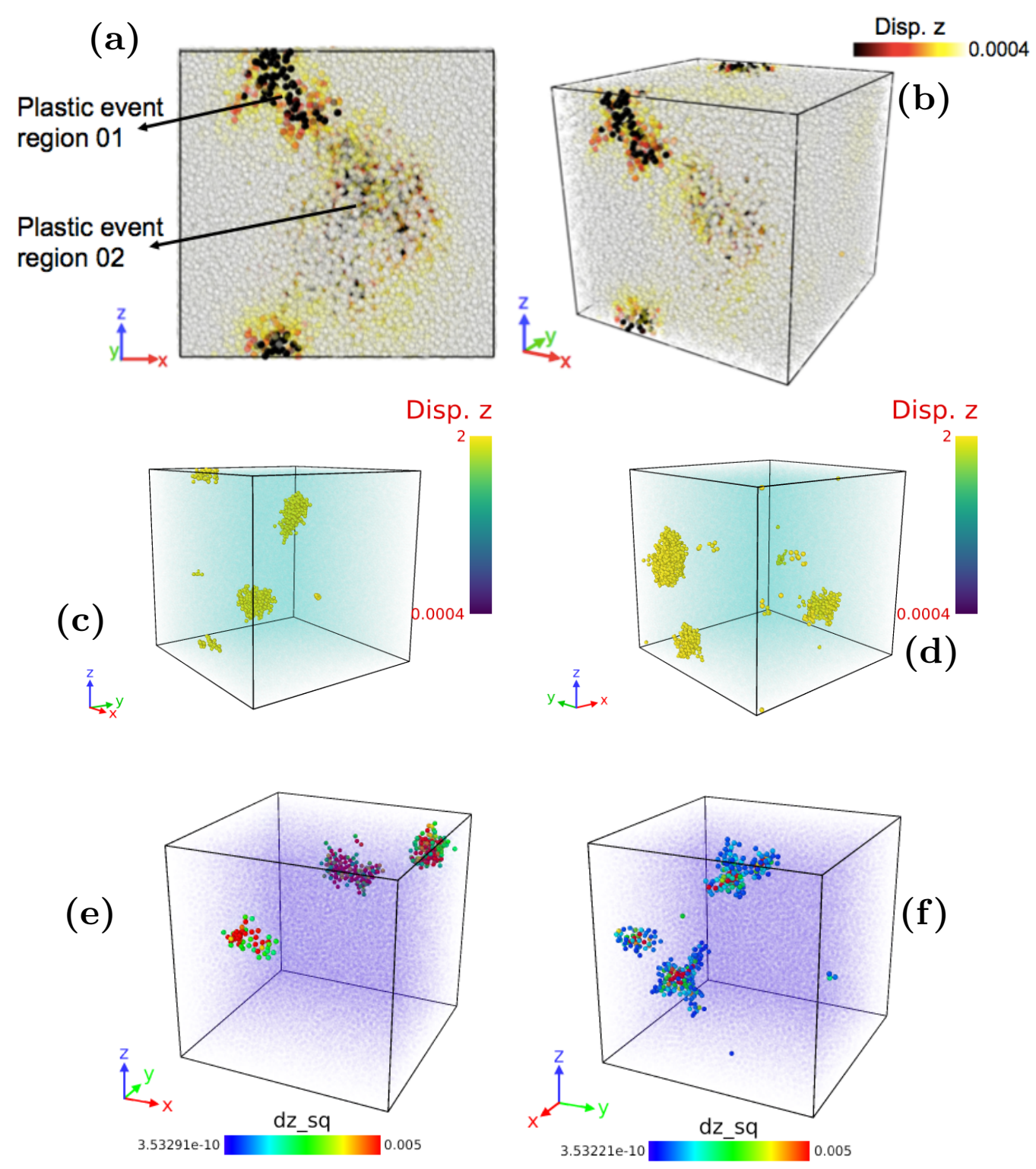

FIG. 1. **Observation of unusual *first* plastic events in silica and metallic glass** with poorly annealed samples. Colored map displays the component of squared displacement in the direction ($z$) perpendicular to the applied shear, in one AQS step of strain increment, see sec. II B. **(a)-(b)**: A representative sample of *VSPBKS silica* with system size $N = 46656$ particles from two different perspectives. The local yield strain at the first plastic event is $\gamma_p = 0.3\%$. **(c)-(d)**: Representative samples of *VSPBKS silica* with a larger system size $N = 216000$ particles. Corresponding $\gamma_p \approx 0.09\%$, $0.1\%$ respectively. **(e)-(f)**: Representative samples of $Cu_{64.5}Zr_{35.5}$ *metallic glass* with $N = 50000$ particles and $\gamma_p \approx 0.1\%$, $0.2\%$ respectively. Clearly, in both types of glasses, there are multiple disconnected cores of high displacements indicating plastic events appearing in multiple regions at the same applied strain $\gamma_p$.

This suggests that these type of events should be quite common in oxide and metallic glasses.

*a.* ***Cell based cluster analysis*** In order to quantify whether the first plastic event in real space is really localized or not, we identify particles that have undergone high displacements in a single strain increment step, and denote them as "plastic particles". At a plastic event, such particles are expected to form cluster(s). In order to count whether the number of cluster(s) is more than one, we employ a cell-based, coarse-grained cluster analysis. Fig. 2(a) illustrates the procedure. We divide the simulation box by grids into $N_c$ smaller cells with $n_x, n_y, n_z$ being the number of cells in $x, y, z$ directions respectively, *i.e.* $N_c = n_x\, n_y\, n_z$. Suppose there are two distinct plastic event regions highlighted in (i) red and

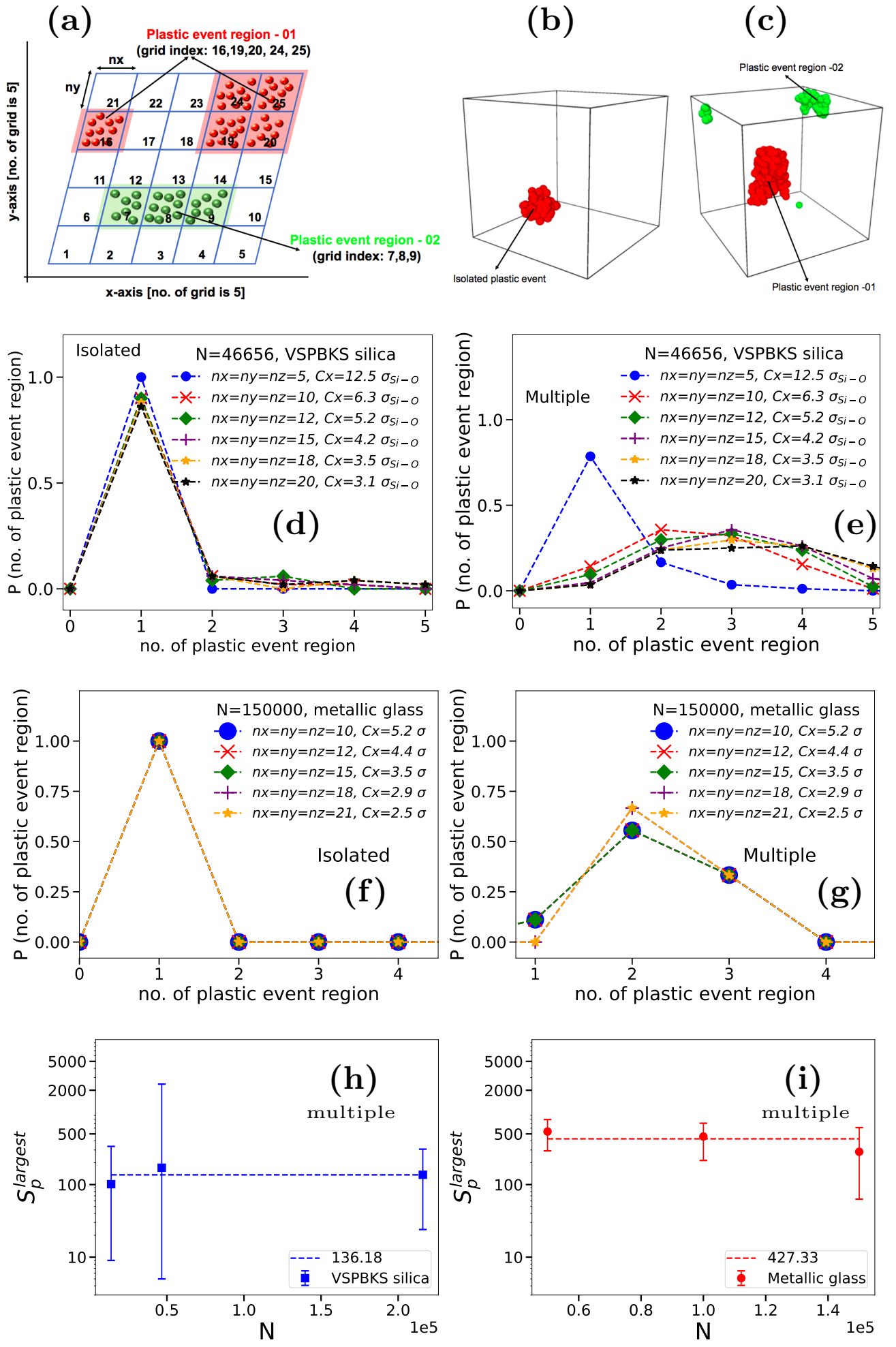

FIG. 2. **Cell based cluster analysis to determine the number and size of distinct plastic cores**. **(a)**: Schematic illustration of the cell-based cluster analysis in real space in a sample. Contiguous colored cells contains a cluster of "plastic particles", while the white cells contains at most one such particle. **(b),(c)**: Classification of the first plastic event in the shear startup regime as (b) "isolated" or (c) "multiple"-type using the cell-based cluster analysis with grid length $4.2\sigma_{Si-o}$. **(d),(e)**: Distribution of number of distinct plastic event regions in *VSPBKS silica* for samples showing (d) **isolated** and (e) **multiple**-type first plastic events in real space. Both types of events are shown for quenched, *i.e.*, poorly annealed samples but at the same degree of annealing. **(f),(g)**: Corresponding distributions in $Cu_{64.5}Zr_{35.5}$ *metallic glass*. We vary the cell size to control resolution and show that there exists a range of resolution over which the distribution is not sensitive to the choice of cell size. **(h),(i):** System size dependence of the size of the largest cluster of plastic particles ($S_p^{largest}$) for (h) *VSPBKS silica* and (i) *metallic glass* respectively. This cluster defines the size of the *core* of a "multiple"-type *first* plastic event. In both materials, changing the system size by an order of magnitude does not significantly change the size of the plastic core.

(ii) green. The red region contains a cluster of plastic particles spanning cells 16, 19, 20, 24, and 25, while the green region shows a cluster contained in cells 7,8, and 9. A cell containing at most one plastic particle is represented in white. In this way we coarse-grain the fractal shape of the cluster by an enclosing polyhedron of connected cells. Clusters of plastic particles (colored cells) thus detected are considered to be disconnected if they are separated by at least one layer of white cells having a cell length in $x$ direction as $C_x = L_x/n_x$ where $L_x$ is the simulation box length along $x$ direction, and similarly for $y, z$ directions. Note that the input configurations (first plastic events) are classified as "isolated" (Fig.2(b)) or "multiple" (Fig.2(c)) by visualization and then the cluster analysis is done on these distinct datasets. In this method, the resolution, *i.e.* the size of a single cell may affect the counting. Hence we systematically vary the cell dimension $C_x$ (or equivalently $n_x, n_y, n_z$).

In Figs. 2(d),(e) we plot the cluster number distribution for the "isolated" and "multiple" type events respectively for representative, poorly annealed samples of VSPBKS silica. As expected for "isolated" events, the peak of distribution robustly occur at one for all $C_x$. In contrast, for the "multiple" type events, for the largest cell-size, the peak occurs at 1, suggesting the cell size to be bigger than typical separation between two clusters. However, as $C_x$ reduces, the peak of the distribution clearly shifts towards higher values and over a broad range of cell dimension, peaks at three. It is somewhat surprising that the "multiple" type events show more than two clusters. However, note that in Figs. 2(d),(e) both types of events occur at the *same* degree of annealing. From these plots, we have chosen the grid length $C_x = 4.2\,\sigma_{SiO}$ as the default value for VSPBKS silica for further analyses. Figs. 2(f),(g) compares the distribution of cluster numbers in metallic glass. Note that for metallic glass, the peak occurs at 2.

To determine whether the "multiple"-type *first* plastic events thus identified are localized or system spanning, we estimate the size of the *core* of an event. Using the cell-based cluster analysis method we determine the size of the largest cluster ($S_p^{largest}$) of plastic particles as a function of the system size $N$. Fig. 2(h),(i) shows the results for VSPBKS silica and metallic glass respectively. The size of the largest cluster does not change significantly when the system size is varied by an order of magnitude. Hence we infer that the "multiple"-type events are localized in space, in accordance with the conventional expectation. However, we note that the number of plastic particles thus identified are $\mathcal{O}(10^2)$ and somewhat larger than what is typically found for LJ-type glasses. This may be due to the longer range of interaction in case of silica and metallic glass, see Fig. 8.

*b.* ***Reproducibility of multiple-type first plastic events*** To establish that the "multiple"-type events are not mere artefacts of the numerical procedure, we check whether the events are reproducible and robust against variation of control parameters of the simulation. In Fig.

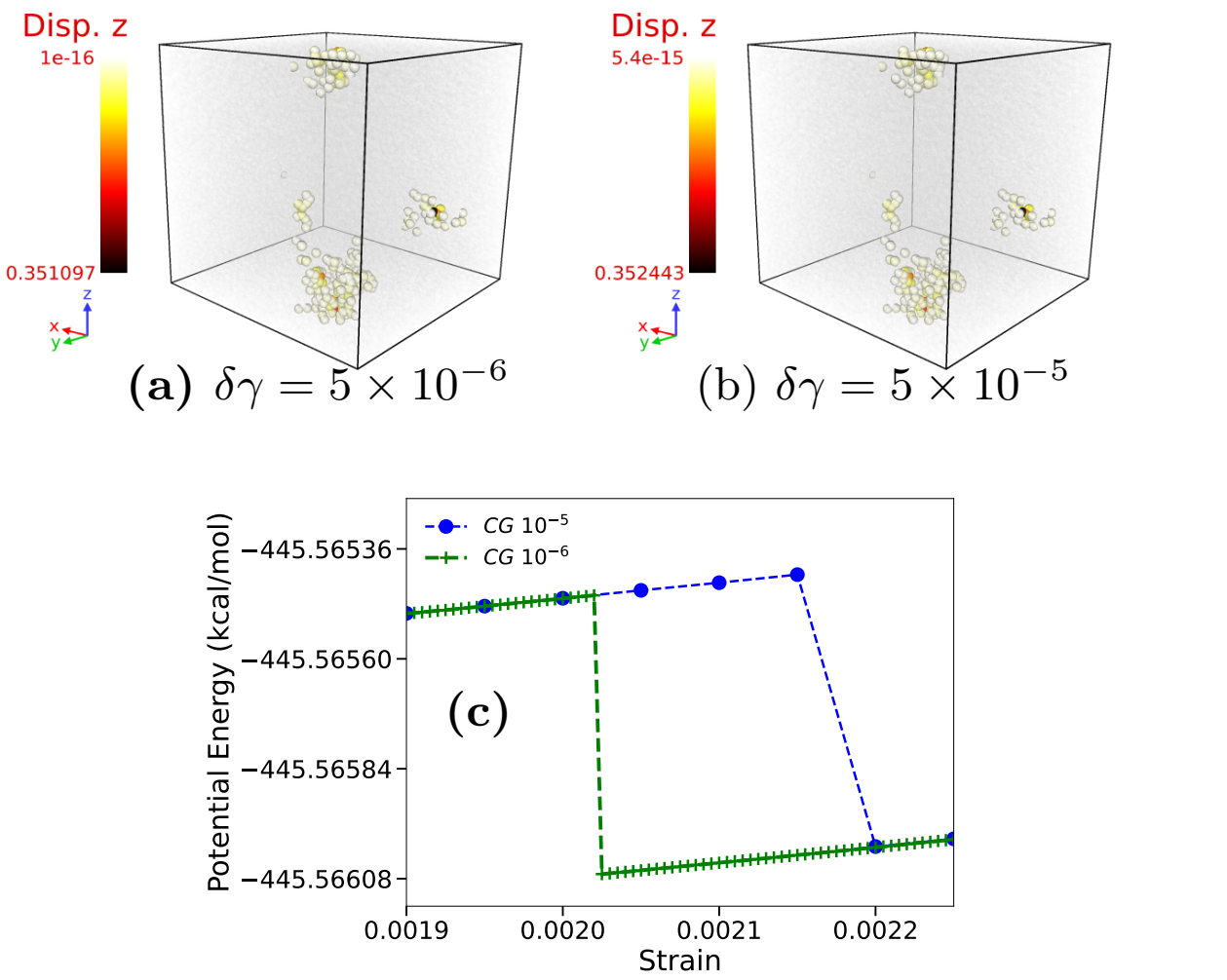

FIG. 3. **Effect of varying AQS strain increment** $\delta\gamma$ using CG minimization, with **(a)** $\delta\gamma = 5 \times 10^{-6}$ and **(b)** $\delta\gamma = 5 \times 10^{-5}$. Only particles displacements ($z$ components of incremental MSD) exceeding a threshold (and thus should be involved in the plastic event) are shown. The color bar is based on the magnitude of the displacement. **(c):** Per particle potential energy *vs.* strain showing discontinuity at the first plastic event. "Multiple"-type events are observed even for a smaller $\delta\gamma$. Samples from *VSPBKS silica* glass.

3(a-b), we compare visualizations of the first plastic event in representative sample of VSPBKS silica when the AQS strain increment $\delta\gamma$ is reduced by an order of magnitude, *i.e.*, from $5 \times 10^{-5}$ to $5 \times 10^{-6}$. Only particles with displacements ($z$ components of incremental MSD) exceeding a threshold (and thus should be be involved in the plastic event) are shown. We see that the nature and real-space location of the first plastic event remains unchanged. Fig. 3(c) shows the per-particle potential energy vs. applied strain starting from the same configuration. Note that with finer resolution, the local yield strain values are slightly lower, implying that the actual path taken during an AQS step in the underlying potential energy landscape somewhat depends on $\delta\gamma$. We have verified that the metallic glass also exhibit the same behaviour.

## B. Cascade nature of "multiple" type first events

Do the distinct spatial regions of plastic activity occurring at the same AQS strain also appear simultaneously in *time*? To answer this question we analyze the relaxation dynamics *during* the (first) plastic event. After applying affine deformation at the local yield strain $\gamma_p$, we turn off the energy minimization and switch to Langevin dynamics (LD). Figs. 4(a)-(c) show the time evolution of per particle potential energy (PE), per particle kinetic energy (KE), and stress for a representative sample of VSPBKS silica for $N = 46656$ particles, along with the corresponding values at the end of the first drop obtained by AQS CG minimization (horizontal dash lines). Each spike in KE corresponds to a sharp drop in PE. We observe multiple sharp drops (spikes) in PE and stress (KE). Note that this does not necessarily imply that the event has a cascade nature, *i.e.* in principle even an "isolated"-type plastic event can show similar relaxation profile. However, we show a sequence of snapshots of increment in transverse (perpendicular to imposed shear) component of MSD: $\delta z_i^2(t+\Delta\, t) = [z_i(t+\Delta t) - z_i(t)]^2$ for each particle $i$ in Figs. 4 (d)-(f) which depicts the progression of the stress relaxation and clearly shows emergence of multiple disconnected clusters one after another. The cascade nature is more evident in Figs. 4 (g)-(i) using a *bigger* system size of $N = 216000$ particles for which we typically find a larger number of distinct plastic clusters. The "multiple"-type events observed in the metallic glass also show the same cascade nature, as demonstrated for a representative sample of system size $N = 50000$ in Figs. 5(a)-(f). Note that in this case, each peak corresponds to a different cluster, and they appear and also disappear at different times.

Thus the physical hypothesis that emerges from our analysis is that multiple "soft spots" in the system can have similar stability against shear deformation and one local shear transformation can trigger a second event *via* the elasticity of the surrounding medium, as recently found in Ref.[26]. Consequently, the non-Affine displacement field at the first plastic event becomes relatively more extended in nature. The surprising fact that this can occur at a much *smaller* deformation than in typical LJ-type glasses, implies that oxide and metallic glasses are perhaps *more* ductile compared to simple, colloidal glasses. To test this somewhat counter-intuitive conjecture, we tune several factors that are known to affect ductility (brittleness) of a glass. The results are presented in the next few sections.

## C. Effect of tuning degree of annealing of initial undeformed sample

It is well known that properties of amorphous solids are influenced by the sample preparation history *e.g.* the starting temperature and pressure (density) of the parent liquid, the cooling rate as they fall out of equilibrium, age of the sample *etc.*[3,4]. In particular, the yielding behavior of a poorly annealed sample show ductile nature and is phenomenologically distinguishable from that of a well-annealed sample exhibiting brittleness[3]. Since more brittle glasses have higher degree of strain localization[31,42], it is interesting to understand how the degree of annealing of the initial, undeformed sample affects the nature of the first plastic event.

Analyses in the previous sections use quenched samples *i.e.* prepared at infinite cooling rate. They are poorly annealed, by definition. Now we prepare samples (VSPBKS silica, $N = 46656$) with varying degree of *finite* cooling rates and determine the nature of the observed *first* plastic events - whether "isolated" and "multiple"-type using the cell-based cluster analysis, sec. III A. Fig. 6(a) shows the dependence of the fraction of each type of events on the cooling rate (lower cooling rate implying better annealed) computed from 10 independent samples. The fastest cooling rate ($10^{-1} K/fs$) shows greater proportion of samples (8 out of 10) showing "multiple"-type *first* plastic events. However, as the cooling rate *decreases*, the fraction of "multiple"-type events *decreases*. Although we note that even for the slowest cooling rate feasible ($10^{-3} K/fs$) 2 out of 10 samples shows "multiple"-type first plastic event, *i.e.*, in VSPBKS silica even in well-annealed samples, the probability of "multiple"-type plastic events remains significant. Fig. 6(b) shows the distribution of number of plastic cores as a function of cooling rate. We see that the width of the distribution systematically decreases and the peak shifts from two to one as the cooling rate decreases. Thus Fig. 6 shows that increasing the degree of annealing makes the first plastic event more localized in space, consistent with the trend that better annealed samples are more brittle and have higher degree of strain localization[31,42].

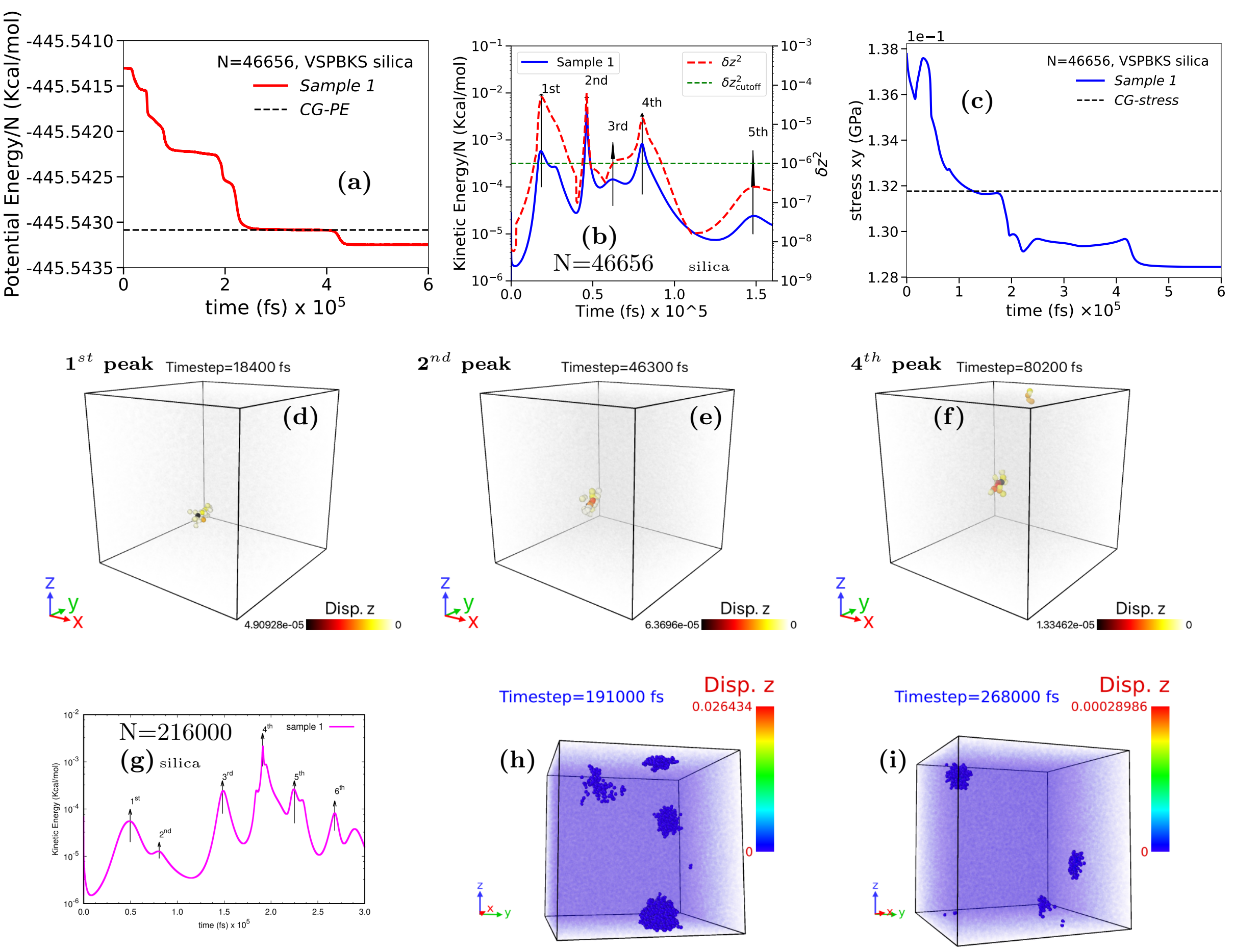

FIG. 4. **Relaxation dynamics during a "multiple"-type first plastic event** for a representative sample of *VSPBKS silica* monitored by **(a)** per particle potential energy, **(b)** per particle kinetic energy, and **(c)** stress (XY component) obtained by Langevin dynamics simulation for $N = 46656$. Also shown in **(b)**, is the mean transverse component of incremental squared displacement $\delta\, z^2(t) = \sum_{i=1}^{N} [z_i(t_0 + t) - z_i(t_0)]^2$ where the reference times $t_0$ were taken approximately at the minima preceding the chosen peaks. **(d)-(f):** Sequence of snapshots showing time evolution of the transverse component of incremental squared displacement for those particles $i$ whose displacement $\delta z_i^2$ exceed the threshold value $\delta\, z^2_{cutoff}$ marked in panel **(b)**. Panels (d),(e) shows one cluster while a second cluster appears in panel (f). **(g)**: Per particle kinetic energy profile for a higher system size $N = 216000$. **(h),(i)**: Sequence of snapshots of $\delta z^2$ field showing appearence of clusters at multiple locations ($N = 216000$). For clarity, the background particles (not participating in the first plastic event) are made transparent in all the snapshots. Note that the core of the plastic cluster with highest displacement is hidden.

### D. Effect of modifying inter-particle interaction

The inter-particle interaction or the force-field is a microsopic control factor which can be tuned to gain insights about brittle-to-ductile transition[31]. Many differ-ent aspects of the force-field, however, can affect the brittleness (ductility). Here we modify two features of the inter-particle interaction and test the effects on the nature of the first plastic event.

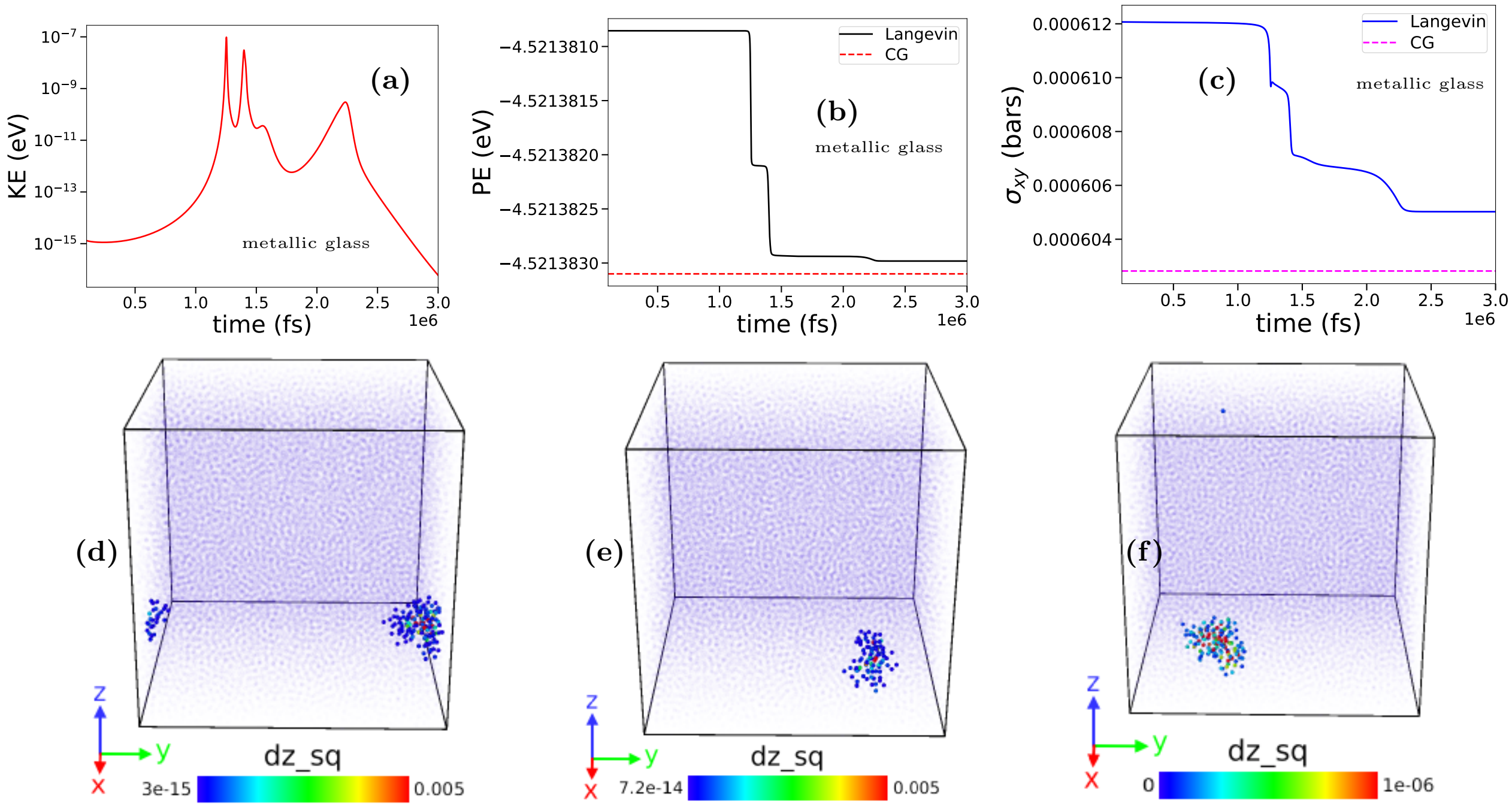

FIG. 5. **Relaxation dynamics during a "multiple"-type first plastic event** for a representative sample of *metallic glass* monitored by **(a)** per particle kinetic energy, **(b)** per particle potential energy, and **(c)** stress (XY component) obtained by Langevin dynamics simulation for $N = 50000$. **(d)-(f):** Sequence of snapshots at 1st ($t = 1.26 \times 10^6$ fs), 2nd ($t = 1.41 \times 10^6$ fs) and 3rd peak ($t = 2.2 \times 10^6$ fs) during Langevin dynamics showing time evolution of the transverse component of incremental squared displacement. Particles with large (non-Affine) displacements ($\delta z_i^2$) are highlighted. Clearly, different clusters seen at the same AQS strain are appearing (and disappeaaring) at different times during LD.

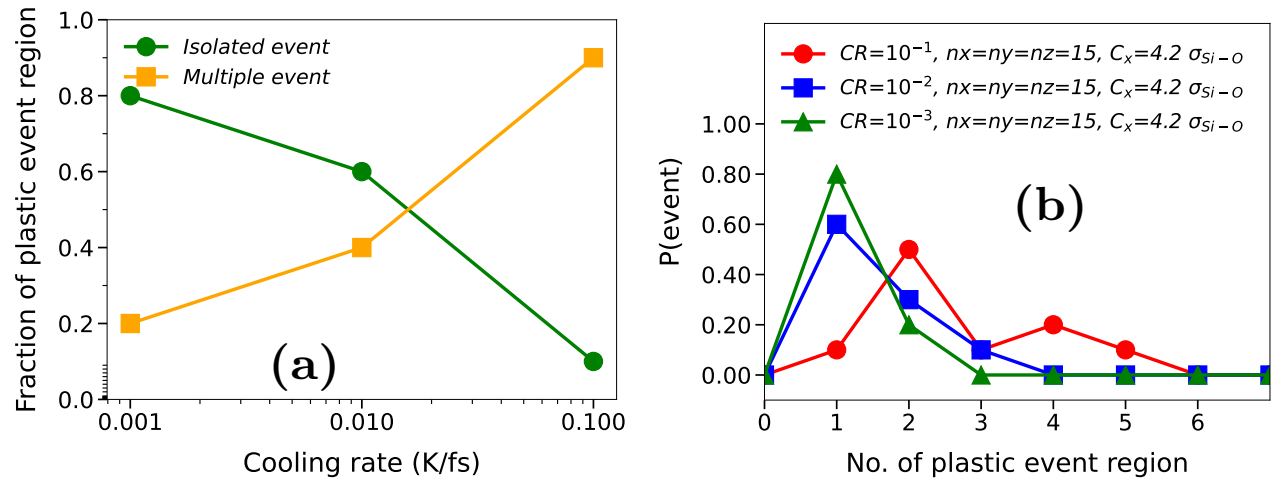

FIG. 6. **Effect of the degree of annealing of the initial, undeformed sample** on the occurrence of cascade-type events. (a) The fraction of isolated and multiple-type events as a function of cooling rate during initial sample preparation. (b) The distribution of number of distinct clusters in the samples including both "isolated" and "multiple"-type first plastic events. With higher degree of annealing *i.e.* slower cooling rate, the peak shifts from two to one and the distribution becomes narrower. Samples from *VSPBKS silica*, $N = 46656$.

*a.* ***Covalent bonded interaction*** The inter-atomic interactions in silica is complex with diverse physical origins such as Coulomb interaction between ions, van der Waals type dipolar interaction as well as covalent bonding to form tetrahedral network[35,43]. Ref.[44] showed that in silicon - a network-former similar to silica - making the covalent bonded interaction *stronger*, *increases* brittleness. Ref.[44] modelled the covalent bonded interaction by a 3-body bond-angle potential, which is however absent in VSPBKS silica force-field. Hence we next choose a second model of silica, *viz.* the Vashihsta 1990 potential[38] which *explicitly* describes the covalent bonding *via* a 3-body bond-angle interaction.

Fig. 7 summarizes the effect of "switching on" the 3-body interaction in silica. We show observation of "multiple"-type *first* plastic event in a representative quenched sample of Vashishta silica in Fig. 7(a). Note that the local yield strain values ($\gamma_p$) are in the same range for both the models of silica studied here. However, for the Vashishta model, we find that the fraction of samples showing "multiple"-type events is significantly *less* than that for the VSPBKS silica. In Fig. 7(b),(c) we quantify the cluster number distribution for "isolated" and "multiple"-type events respectively and in Fig. 7(d) we show the $N$ dependence of the size of the large cluster of plastic particles (all quenched samples). While the profile of "isolated" type events are similar in both models (*cf:* Fig. 2), the peak of the "multiple"-type events in Vashishta silica occur at two even for the quenched samples.

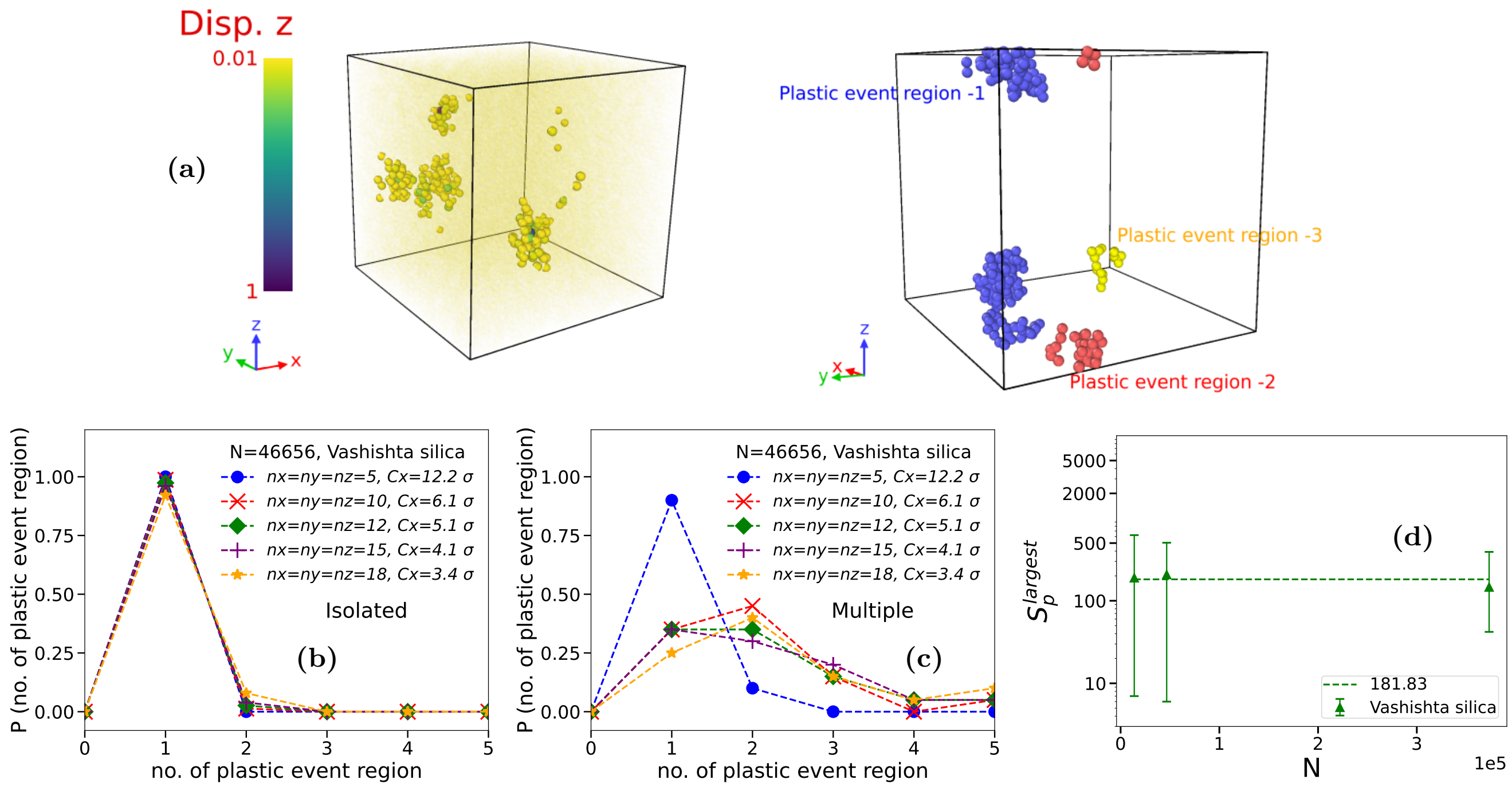

FIG. 7. **Observation of "multiple"-type *first* plastic events in the Vashishta silica model. (a):** Real space visualization of particles undergoing high displacements perpendicular to applied shear ("plastic" particles) at the local yield strain for a representative sample with $\gamma_p \sim 0.27\%$. Multiple distinct *cores* of shear transformation is clearly visible. For clarity particles exceeding a threhold value of local maximum displacement are highlighted. For the Vashishta silica, 28 out of 100 samples (all quenched) appear to have "multiple"-type first events upon visualization. **(b)-(c):** The cluster number distribution for samples showing **(b)** "isolated" and **(c)** "multiple"-type first plastic events respectively, computed by the cell based cluster analysis. The contrasting profiles highlights the distinction between them. Note also that the peak for "multiple"-type events for Vashishta silica occur at two, even for quenched samples, unlike that of VSPBKS silica. This implies that Vashsihta silica model is *less ductile* than VSPBKS silica. **(d)**: System size dependence of the size of the plastic core as measured by the size of the largst cluster of plastic particles $S^p_{largest}$.

Thus we demonstrated the observation of "multiple"-type events using two different silica force-fields. Further, we find that the extra 3-body covalent interaction in Vashishta silica makes the system *less* ductile than VSPBKS silica, which is consistent with the exisitng literature[31,44].

*b.* ***Range of interaction*** The analyses in the previous sections suggest that this type of event is not a feature of a specific model potential, but a *generic* physical property of a broad family of glasses. The question naturally arises about why such "multiple"-type first plastic events are more frequent in silica and metallic glass but are not well known in the more commonly studied LJ type glasses. Note that both the silica models studied here have (medium) long ranged (screened) Coulomb interaction. Consequently the range of interaction for both VSPBKS and Vashsihta silica is longer than typical LJ type glasses. The interaction cutoffs are $r_c = 10\mathring{A} \approx 7.6\sigma_{SiO}$ and $r_c = 10\mathring{A} \approx 7.7\sigma$ for VSPBKS and Vashisht silica models respectively. Increasing the range of interaction is known to make a glass more ductile[45]. Hence we now consider the long-ranged Kob-Andersen (LKA) binary mixture and study the effect of increasing the range of interaction.

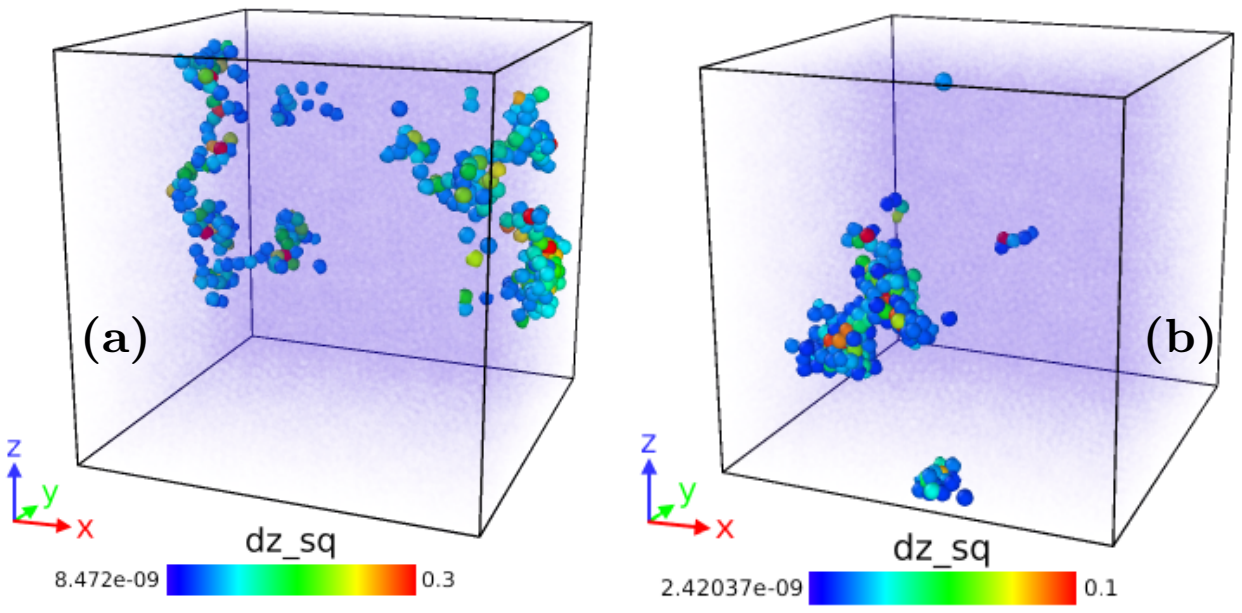

FIG. 8. **Effect of increasing the range of interaction in Kob-Andersen glass.** Transverse squared displacement field at the local yield strain $\gamma_p$. For clarity, (plastic) particles with large non-Affine displacements are highlighted. **(a):** "Isolated"-type first plastic event has large cluster size. **(b)** Observation of "multiple"-type event (2 out of 15 samples). Taken together, they suggest that increasing range of interaction make the system *more* ductile.

Fig. 8 shows snapshots of transverse component of squared displacement fields at the first plastic event (at $\gamma_p$) for representative samples of a KA type glass with longer cutoff. The "isolated" type events [panel (a)] shows larger size of the plastic *core*. In addition, 2 out of 15 samples analyzed now show "multiple"-type events [panel (b)]. Taken together, they clearly show that the range of inter-particle interaction affects the nature of non-Affine displacement field. Increasing the range of interaction makes the plastic displacement less localized in space, consistent with earlier studies[45].

### E. Effect of tuning background stiffness by "soft matrix" method

*a.* ***Soft matrix method (SM)*** The stress relaxation dynamics in Sec. III B indicates that even at the *first* plastic event, the plastic instabilities proceed in a cascade-like fashion, where a localized plastic event can trigger a subsequent event in a second region, even at the same applied strain (AQS). In order to test this picture further, we now anayze the first plastic events using a probe of local mechanical response recently developed by some of us and termed as the "soft matrix" method, see Ref.[33] for details. In this method, the system is divided into two parts - (i) *target region:* a local sub-box of dimension $L_s$ where the particle displacement is unrestricted and (ii) *background matrix:* the surrounding region where non-Affine (plastic) displacement is hindered by attaching springs to individual particles. The spring constant $k$ represents an elastic stiffness of the background matrix in addition to the inherent matrial elasticity. Although in principle it can be spatially heterogenerous, for simplicity we choose uniform value of $k$ for the background. Inside the target region sub-box, $k = 0$.

*b.* ***Role of background elasticity*** In the SM approach, the spring constant $k$ can be tuned to control the stiffness of the background matrix. To this end, we analyze *first* plastic event with two (and not more) distinct cores as detected by direct visualization and cell based cluster analysis in unconstrained AQS. We define a sub-box of dimension $L_s$ around one of the two distinct plastic cores as the target region. Then we perform AQS starting from the same initial undeformed sample but for varying $k$, *i.e.* for different stiffness of the background matrix. Figs. 9(a)-(c) and 9(d)-(f) present the results for (quenched samples of) the VSPBKS silica and the metallic glass respectively. For this analysis, the dimensions of sub-box are chosen to be $L_s \approx 28\sigma_{SiO}$ and $L_s \approx 25\mathring{A}$ for VSPBKS silica and metallic glass respectively. These values are much smaller than the dimensions of the simulation boxes $L \approx 82\sigma_{SiO}$ for the VSPBKS silica and $L \approx 93\mathring{A}$ for the metallic glass respectively. Fig. 9(a),(d) show the number of distinct clusters (plastic cores) of the *first* plastic events as $k$ is gradually increased by several orders of magnitude. Although the absolute values of $k$ are different in silica and metallic glass, due to the intrinsic difference in the interaction energy scales in the two systems, both of them show the same qualitative behaviour. The number of distinct cores remains same (=two) as the unconstrained ($k = 0$ everywhere) AQS upto a *finite* value of $k$. It is also verified that the real space location of the plastic events remains approximately the same for zero and fintie $k$ values. However, as $k$ is increased further, only one plastic core (inside the target region) is observed at the first plastic event. The contrast in behaviour is further clarified by plotting the strain ($\gamma$) dependence of the component of MSD transverse to the imposed shear at two different $k$ values, in Figs. 9(b)-(c) and 9((e)-(f) respectively for the VSPBKS silica and the metallic glass. Here the data-sets "Full" and "L01", "L02" refer to the unconstrained AQS and the target regions during SM AQS respectively. When the background is sufficiently stiff, the plastic event outside the target region is sufficiently delayed so that only the event inside the target region occurs at the local yield strain. Thus the imposed background elasticity acts as a barrier to hinder plastic displacements. This analysis suggests that tuning the background elasticity may be an interesting route to understand and engineer brittle-to-ductile transition in amorphous solids.

## IV. SUMMARY AND CONCLUSIONS

In the present work we study shear deformation response of a prototype oxide glass *viz.* silica using two different force fields, as well as a model $Cu_{64.5}Zr_{35.5}$ metallic glass, in the transient, elastic, shear start-up regime by athermal, quasistatic simulation. We robustly show that even the *first* shear transformation (plastic) event - at applied strain two orders of magnitude smaller than the yield strain - can have a complex real-space structure with *multiple disconnected cores.* By monitoring stress relaxation dynamics *during* the first plastic event, we demonstrate the cascade-nature of such events. According to conventional wisdom, such cascade nature should occur near or post yielding steady state, but should be highly improbable at the local yield strain. To gain insight about the nature of the events, we tune three factors that affect ductility ( brittleness) of glass - (a) the degree of annealing of the initial undeformed sample, (b) the inter-particle interaction *viz.* covalent bonding and range of interaction and (c) the stiffness of the (elastic) medium. In all cases we find that factors increasing brittleness decreases the likelihood of such "multiple"-type *first* plastic events. We have also explored whether the nature of first plastic events are different due to different types of bonding between metallic glass and silica. Our preliminary analysis indicates that the differences are quantitative, rather than qualitative, see AppendixB.

While many such high level parameters controlling brittleness have been empirically identified and discussed in the literature[31,44], for physical understanding, *micro-*

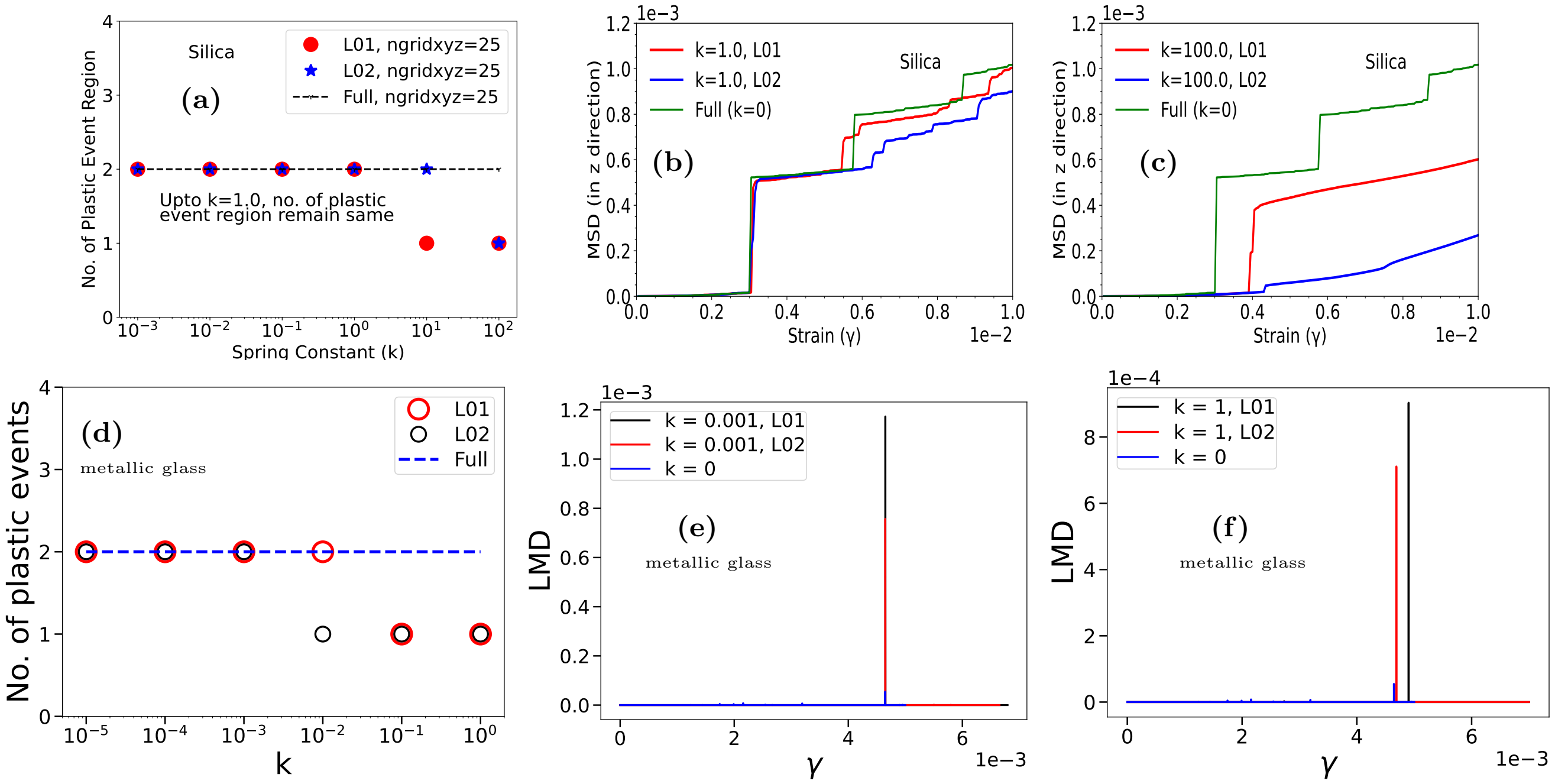

FIG. 9. **Soft matrix (SM) method analysis of the first plastic event** in **(a)-(c)** *VSPBKS silica* and **(d)-(f)** *metallic glass* for samples showing two, well-separated clusters in unconstrained AQS ($k = 0$). Data-sets "Full" and "L01", "L02" refer to the unconstrained AQS and the target regions during SM AQS respectively. **(a):** Number of distinct plastic clusters as a function of background elastic stiffness $k$ when a single target region is probed. **(b),(c):** Transverse component of MSD at two different $k$ values highlighting the contrasting behaviour and showing that increasing the background stiffness can delay the event in the background, such that local yield strain at two regions becomes different from the unconstrained AQS. **(d)-(f):** Effect of varying background stiffness $k$ in *metallic glass*. (d) Number of distinct plastic clusters as a function of background elastic stiffness $k$. (e),(f): Local maxmimum displacement at two different $k$ values displaying the contrast in behaviour. Even in metallic glass, the same behaviour is observed, *i.e.*, increasing $k$ lifts the degeneracy of local yield strain at the two different regions in space.

*scopic* quantifiers of the degree of brittleness of a glass are necessary but largely missing. Several recent studies have explored the intuitive idea that local structural environment should determine the local yield strain and sought to find the relationship among structural defect, softness and avalanches[13,31,44,46,47]. In particular, Ref.[31] showed that degree of local "softness" is a crucial microscopic factor that affect ductility (brittleness). Using a schematic meso-scale model, these authors have shown that increasing the "softness" increases the ductility of glass. Using a recently developed "soft matrix" method, we directly examine the role of the stiffness (local elastic constant) of the elastic background and arrive at the same conclusion. Indeed, our results of varying all macroscopic factors in the present study are consistent with the intuitively appealing picture that "softer is more ductile". Thus our study highlights that understanding the role of the softness field is a promising strategy to quantify brittle-to-ductile transition at microscopic level. This is the main implication of the present work. We however note that the relation between softness and ductility is a topic of ongoing research. Generally it is thought that ductility is related to the real space distribution of softness[3] which may in turn be connected to the topography of underlying PEL[48]. However we note that at higher applied strain approaching yielding point, the nature of mechanical stability may change[49].

A second point to note is that shear deformation response at small strain $\gamma \sim 0.1 - 0.3\%$ actually provides valuable information about the *undeformed* glass and its underlying potential energy landscape (PEL). Indeed, a motivation of exploring the so-called "elastic avalanches" in the recent literature is to understand "marginal stability" related to the hierarchial structure of the PEL[20,28]. While such studies have pre-dominatly looked into simple LJ type soft-sphere models (colloidal glass), the present study suggests that it may be easier to prepare marginally stable sample using oxide and metallic glasses. This is perhaps due to the longer range of interaction in such materials, as our results suggest. We however clarify that although the complex real-space structure of the first plastic events is suggestive of marginal stability, one requires further analysis of the statistics of such events and determine the avalanche exponents to rigorously make the connection.

Thus the present study suggests that oxide and metallic glasses show richer behaviour than the conventional view of plasticity in the shear startup regime, and should be interesting to explore not only for engineering application but also fundamental physical understanding about amorphous solids.

## ACKNOWLEDGMENTS

We thank the National Supercomputing Mission (NSM, via grant DST/NSM/R&D HPC Applications/2021/29) for providing the computational facilities "PARAM Yukti" at Jawaharlal Nehru Centre for Advanced Scientific Research, Bengaluru and "PARAM Ganga" at the Indian Institute of Technology Roorkee, implemented by C-DAC, India and supported by MeitY and DST, Government of India. S.S. thanks the Indian Institute of Technology Roorkee for providing support via the Faculty Initiation Grant (PHY/FIG/100804). H.R. acknowledges junior and senior research fellowships from Council of Scientific Research, India. We thank Vishwas V. Vasisht for helpful discussions. S.S. thanks Himangsu Bhaumik for kindly providing the code for the VSPBKS silica model.

## AUTHOR DECLARATION

### Conflict of interest

The authors have no conflicts of interest to disclose.

### Author Contribution

## DATA AVAILABILITY

The data are available from the corresponding author upon reasonable request.

### Appendix A: Plastic particles

Here we provide more details about the definition of "plastic particles" and the method of identifying the core of a plastic event. Figs. 10(a) and (b) shows representative AQS runs of Vashisht silica ($N = 46656$) during which several indicators are monitored to identify a plastic event. In particular the observable "local maximum displacement" (LMD) $\delta z^2_{max}$ (see II B) is compared against (a) MSD ($z$ component perpendicualr to applied shear) and (b) per particle potential energy (PE). We see that the discontinuities in MSD and PE are accompanied by spikes in LMD. Thus we use LMD to identify a plastic event and measure the corresponding local yield strain $\gamma_p$. To estimate the core of a shear transformation (plastic event) thus identified at $\gamma_p$, particles with large LMD values above a threshold is labelled as plastic particles. The threshold is chosen by plotting the cumulative distribution function $CDF(y) = \int_0^y dy'\, P(y')$ where $y = \delta z^2_{max}$ and $P(y)$ is the probability denstiy function of observing an LMD value in the range $[y, y + dy]$ at *any* strain step during an AQS run. Fig. 10(c) compares the CDF for both the VSPBKS and the Vashishta silica models. Following Ref.[33] the threshold for the VSPBKS silica is chosen to be 0.0004 while that for the Vashishta silica is chosen to be 0.01 in reduced units. The threshold $\delta z^2_{max}$ and the corresponding CDF values are indicated by vertical and horizantal lines respectively. The threshold for metallic glass is $10^{-5}$ (reduced units). Further, we rank the plastic particles thus identified according to their LMD values, and chose the top $x\%$ particles with largest (non-Affine) displacement to identify the *core* of a shear transformation event. The fraction $x$ is chosen such that the core is sufficiently well-defined to make a clear distinction between "single" and "multiple" type events.

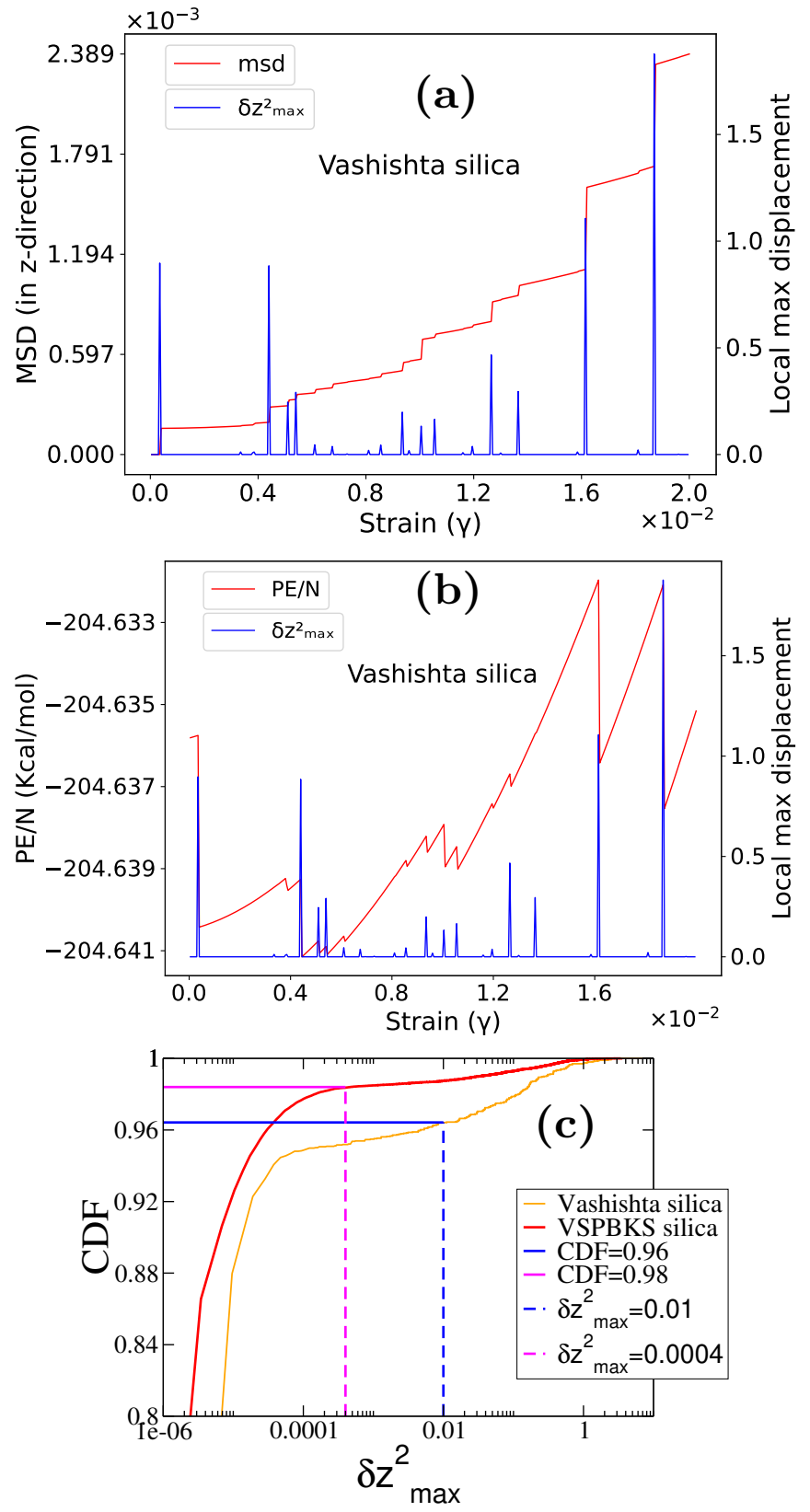

FIG. 10. **(a),(b):** Benchmarking local maximum displacement (LMD) $\delta z^2_{max}$ perpendicular to applied shear as a robust indicator of plastic events by comparing against other standard indicators such as dicontinuities in (a) mean squared displacement MSD (component normal to applied shear) and (b) per particle potential energy using data for Vashishta silica. **(c):** Comparison of the cumulative distribution function (CDF) of $\delta z^2_{max}$ between VSPBKS and Vashishta silica models highlighitng the difference between the two force-fields. Vertical lines indicate the threshold displacement to identify the particles with large displacements at a plastic event and thus identify the core of a plastic event. The values are 0.00004 for VSPBKS silica and 0.01 for Vashishta silica in appropriate units. The horizontal lines indicate the value of the CDF at the threshold values.

## Appendix B: Effect of different bonding types: silica *vs.* MG

Here we analyze the possible difference in the nature of first plastic events between metallic glass and silica due to different nature of inter-particle interaction. Our analysis shows that the differences are mainly quantitative with qualitative features largely similar. Note that in the present study we mostly use quenched samples for silica and moderately annealed samples fo metallic glass. This difference in the degree of annealing makes direct comparison between the two glasses somewhat difficult. Nevertheless, we list the key quantitative differences. (a) First, we note that energy release in the first plastic event during AQS is typically larger for silica than MG, see Fig. 11(a)-(c). Here both "isolated" and "multiple" type events are considered. (b) Similarly, the local maximum displacement (LMD, $\delta z^2_{max}$) values characterizing the particle displacements during the first plastic (considering both types), are larger in silica than MG, as shown by the cumulative distribution function (CDF) in Fig. 11(c). Note that this remains true even if we compare the two glasses as similar degree of annealing. (c) In Fig. 11(d)-(e) we compare the PE and stress relaxation profiles during Langevin dynamics simulations following typical "multiple"-type first plastic events in metallic glass and VSPBKS silica. Thus this difference between the two glasses remains true even if only "multiple"-type first plastic events are considered. (d) Next, there are quantitative differences in the real-space characteristics of the *core* of the "multiple"-type first plastic event in the two glasses. Comparison of cluster number distribution and cluster sizes among Fig. 2 and Fig7 shows that the events are less fragmented and the core size of the largest cluster of plastic particles ($S_p^{largest}$) is bigger in metallic glass compared to silica. Thus, the data shows that a typical first plastic event is "milder" in metallic glass than silica as estimated by energy release and LMD distribution. This is consistent with the expectation that the typical energy scales of metallic bonds in MG are lower than those of covalent bonds in silica. However, the degree of annealing is another crucial factor affecting the properties of plastic events [see *e.g.* the different CDF profiles of VSPBKS silica quenched *vs.* annealed in Fig. 11(c)]. The observed differences in the first plastic event between metallic glass and silica is likely due to several factors such as different nature of bonding as well as different sample preparation history *e.g.* differing cooling rate.

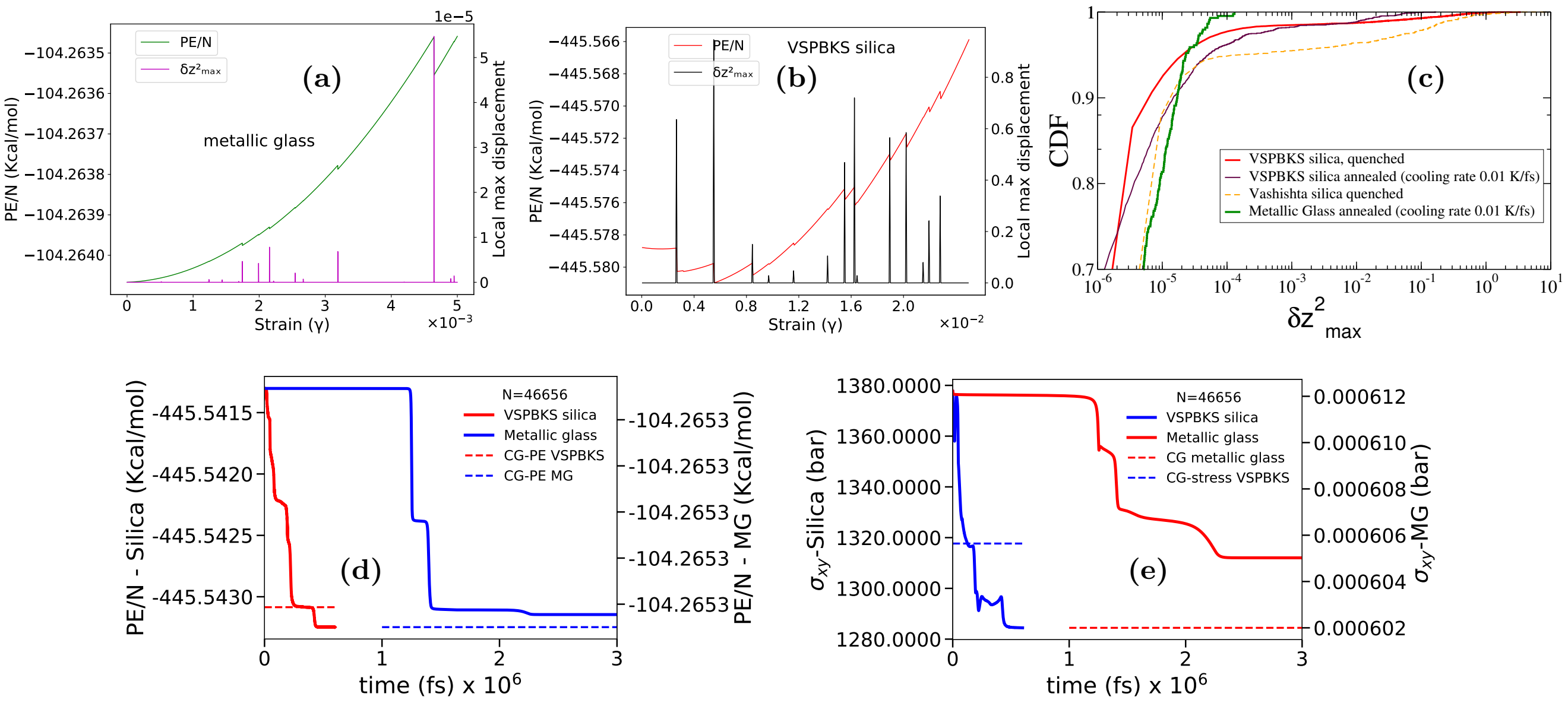

FIG. 11. **(a),(b):** Potential energy (PE) per particle *vs.* strain for metallic glass and VSPBKS silica respectively during AQS runs. Alongside, the local maximum displacement (LMD, $\delta z^2_{max}$) is shown in the right y-axis. Note that the events shown in (a)-(b) are both "isolated" and "multiple" types. **(c):** The cumulative distribution function (CDF) profile of VSPBKS silica (quenched and annealed), Vashishta silica and metallic glass. **(d)-(e):** Comparison of per particle PE and stress relaxation profiles between metallic glass and VSPBKS silica during Langevin dynamics simulation following typical "multiple"-type first plastic events. The horizontal lines of same colors corresponds to their energy minimized value during AQS on their respective axes.